\documentclass[final,1p,times]{elsarticle}

\usepackage{amssymb}
\usepackage{amsmath}

\journal{Nuclear Physics A}

\begin{document}

\begin{frontmatter}

\title{Analysis of the near-side ridge structure in pp collisions via Momentum-Kick Model}

\author[first]{Jaesung Kim}
\ead{Jaesung\_Kim@inha.edu}

\author[first]{Jin-Hee Yoon\corref{cor1}}
\ead{jinyoon@inha.ac.kr}

\cortext[cor1]{Corresponding author}

\affiliation[first]{
            organization={Department of Physics, Inha University},
            addressline={100, Inha-ro}, 
            city={Incheon},
            postcode={22212}, 
            country={Republic of Korea}}

\begin{abstract}
The near-side ridge structure has been observed in the long-range two-particle correlations in heavy-ion collisions, such as AuAu collisions at the Relativistic Heavy Ion Collider(RHIC) and PbPb collisions at the Large Hadron Collider (LHC).
Hydrodynamic models have successfully explained the ridge structure in heavy-ion collisions, indicating the presence of Quark-Gluon Plasma (QGP).
Interestingly, similar ridge structures have been detected in high-multiplicity proton-proton and proton-lead collisions, which are classified as small systems in the LHC experiments.
Because small systems have been considered insufficient to generate QGP, the applicability of theories developed for heavy-ion collisions to small systems remains controversial.
Assuming that kinematic effects play a more significant role in small systems, we expect that the Momentum-Kick Model (MKM) can provide a satisfactory explanation.
This model elucidates the long-range and near-side ridge structure in dihadron $\Delta\eta-\Delta\phi$ correlation by explaining that jet particles kick and rearrange medium partons along the direction of the jets.
In this study, we apply the MKM to explain high-multiplicity proton-proton collisions at both 13 TeV and 7 TeV in the LHC over various ranges of momenta.
Furthermore, we introduce multiplicity dependence in the model to account for the 13 TeV data at various multiplicity ranges.
We conclude that the MKM effectively explains the near-side ridge structure observed in proton-proton collisions.
The LHC has entered Run 3, achieving higher center-of-mass energies and better luminosity than Run 2.
We offer $\Delta\phi$ correlation predictions for pp collisions at 14 TeV and suggest possible extensions of the MKM for future studies.
\end{abstract}





\end{frontmatter}




\section{\label{sec:Introduction}INTRODUCTION}

The ridge structure refers to the shape of the ‘ridge’ in the dihadron $\Delta\eta - \Delta\phi$ correlation, which appears at high $\Delta\eta$ range.
This phenomenon was initially discovered in heavy-ion collisions, such as AuAu collisions at the Relativistic Heavy Ion Collider (RHIC) \cite{ref12, ref13, ref14, ref15, ref16, ref17, ref18, ref19, ref20, ref21, ref22, ref23, ref24, ref25, ref26, ref27} and PbPb collisions at the Large Hadron Collider (LHC) \cite{ref1, ref2, ref3, ref4, ref5, ref6, ref7, ref8, ref9, ref10, ref11}.
Since the medium created in heavy-ion collisions exists in a high-temperature and high-density environment, the ridge structure can be understood through collective motion based on hydrodynamic theory.
This is a hint of the Quark-Gluon Plasma (QGP) state.
Recently, however, observations have shown that similar ridge structures also appear in small systems.
Since small systems were not expected to foster an environment to generate collective motion, researchers have been prompted to investigate alternative explanations beyond hydrodynamic models.

Currently, two mainstream mechanisms have been employed to elucidate the near-side ridge structures in small systems.
These mechanisms include the glasma correlation in the initial state \cite{cgc1, cgc2, cgc3, cgc4, cgc5, cgc6, cgc7, cgc8} detailed by the Color Glass Condensate (CGC) effective field theory and the final state evolution \cite{hydro1, hydro2, hydro3, hydro4, hyro5} described by hydrodynamics.
However, a complete understanding of the ridge structure remains unsolved.
In small systems, due to the smaller number of particles produced compared to heavy-ion collisions, the medium created from collisions can be treated as partons that interact solely through kinematic interactions.
Therefore, we expect that pure kinematic models may offer a plausible explanation for the long-range and near-side ridge structure in small systems.

The Momentum-Kick Model (MKM) is based on purely kinematic interactions between the near-side jets and medium partons \cite{Wong_2, Wong_3, Wong_4, Wong_5}.
Given that jet fragments consist of particles with exceptionally high transverse momentum, it is logical that these jet particles kick the surrounding medium partons, rearranging them along the direction of the jets.
This process is similar to the collective motion observed in heavy-ion collisions.
The MKM elucidates the ridge structure through this process.
Since the MKM has been successfully applied to the experimental results in AuAu at 200 GeV \cite{Wong_1}, PbPb at 2.76 TeV \cite{PbPb}, and high-multiplicity pp at 13 TeV \cite{Hanul}, we need to scrutinize whether the model can effectively explain the ridge structure in small systems.

All LHC experiments are currently investigating the ridge structure in small systems across various environmental conditions, including multiplicity ranges, center-of-mass energies, $p_T$ ranges, etc.
Thus, it should be verified whether the MKM can explain all the data in different environments.
In this paper, we will apply this model to high-multiplicity events at 13 TeV and 7 TeV center-of-mass energies, as well as to various multiplicity ranges at 13 TeV, from all three major collaborations: A Large Ion Collider Experiment (ALICE), Compact Muon Solenoid (CMS), and A Toroidal LHC Apparatus (ATLAS).

We introduce the main concepts of MKM in Section \ref{sec:Momentum-Kick_Model}.
In Section \ref{sec:FITTING_RESULTS}, we employ the MKM to analyze high-multiplicity proton-proton collisions at 13 TeV, utilizing data from ALICE, CMS, and ATLAS altogether \cite{alice, cms, atlas}.
We also examine high-multiplicity pp collisions at 7 TeV from the CMS \cite{cms}.
Moreover, Section \ref{sec:FITTING_RESULTS} is divided into three Subsections: Subsection \ref{subsec:Analysis_Procedure} explains the data analysis in experiments, and Subsection \ref{subsec:pp_collisions_Fitting_results} presents the fitting results of high-multiplicity proton-proton collisions at 13 TeV and 7 TeV.
In addition, we discuss proton-proton collisions at 13 TeV for various multiplicity ranges from the ATLAS experiment in Subsection \ref{subsec:How_about_the_other_multiplicity?}.
Section \ref{sec:Prediction} explains our prediction for the dihadron correlation yield at a higher center-of-mass energy, 14 TeV.

\section{\label{sec:Momentum-Kick_Model}Momentum-Kick Model (MKM)}

\begin{figure}
\centering
\includegraphics[width=10cm]{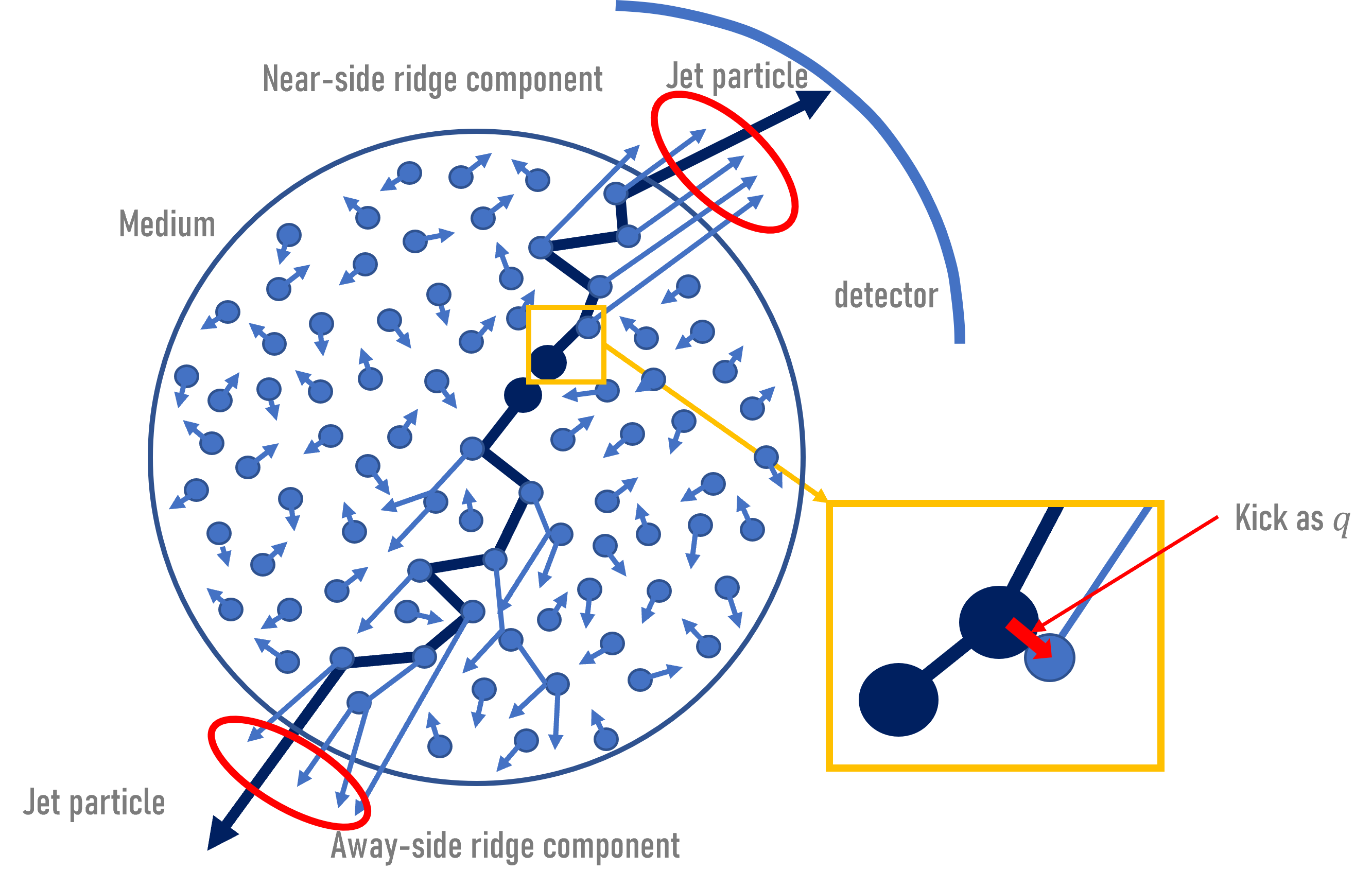}
\caption{\label{figure:description}
Illustration of how the Momentum-Kick Model (MKM) explains the near-side ridge structure.
While the ridge structure includes both near-side and away-side components, the MKM focuses on the near-side ridge by utilizing the idea that jet particles kick medium partons, rearranging them along the jet direction.
}
\end{figure}

As schematically depicted in Figure \ref{figure:description}, the Momentum-Kick Model (MKM) explains the near-side ridge structure by proposing that near-side jet fragments impart transverse momentum to nearby medium partons.
These kicked partons then stream outward along the jet axis, form a collective motion, and hadronize into stable final-state particles. 
In high-energy collisions, pairs of high-energy trigger particles are produced in nearly opposite directions and then interact with the surrounding medium, producing cone-shaped jet fragments.
Since away-side jet fragments undergo multiple scatterings, which makes modeling the individual kicks on the away side difficult; hence, the MKM mainly focuses on the near-side phenomena.
Because small systems create a relatively small number of medium particles compared to heavy-ion systems, we expect that kinematic effects is more dominant in pp or pPb collisions than in PbPb collisions.
Consequently, given that Ref.~\cite{PbPb} has demonstrated the MKM’s success in describing PbPb results, we anticipate that it will be even more effective in describing pp collisions.

In experimental analyses, two particles are randomly selected to generate dihadron correlations.
On the near side, three types of correlation components typically emerge.
The first involves correlations among the near-side jet fragments themselves: since these fragments remain confined in a small cone, correlations predominantly appear at $\Delta \phi \approx0$ and $\Delta \eta \approx 0$.
This component can be largely excluded by using long-range data.
The second component emerges between near-side jet fragments and near-side kicked medium partons: if the initial medium partons are distributed in a plateau shape within mid-rapidity, the near-side jet kicks them, resulting in a peak at $\Delta \phi \approx 0$ across a wide $\Delta \eta$ range.
The third component appears among the near-side kicked medium partons themselves; these kicked partons are propelled outward in a cone with roughly twice the opening angle of the jet cone, which naturally results in a $\Delta \eta$ range nearly twice as broad.
In addition, under the assumption that the initial medium partons form a plateau in mid-rapidity, the third component preserves the shape of the second component.
Thus, the third component does not influence the behavior within mid-rapidity.

To focus on the ridge structure, we select long-range experimental data.
Consequently, we will explain the measurements using the ridge component in the MKM, which represents the yield of medium partons kicked by the jet.
The ridge component in the MKM is represented by the charged ridge particle momentum distribution per trigger jet:
\begin{eqnarray} \label{equation:eq1}
\left[ \frac{1}{N_{\text{trig}}} \frac{dN_{\text{ch}}}{p_Tdp_Td\Delta\eta d\Delta\phi} \right]_{\text{ridge}} = \frac{2}{3} f_R\left\langle N_k\right\rangle \frac{dF}{p_T dp_T d\eta d\phi},
\end{eqnarray}
where $N_{\text{trig}}$ is the number of trigger particles.
The variables $\Delta \eta $ and $\Delta \phi$ represent the differences in pseudorapidity and azimuthal angle between jets and other particles.
The factor $2/3$ denotes the ratio of charged particles to the total number of particles.
The parameter $f_R$ indicates the average survival factor of ridge particles, and $\left\langle N_k\right\rangle$ represents the average number of kicked partons per-trigger particle.
Generally, $f_R$ and $\langle N_k \rangle$ are combined as $f_R \langle N_k \rangle$, and the $p_T$ dependence has been introduced in previous studies \cite{PbPb, Hanul}.

The MKM assumes that kicked medium particles gain a momentum transfer $\mathbf{q}$ from jet particles, which leads to the ridge structure expressed as the final momentum distribution of charged ridge partons:
\begin{eqnarray}\label{equation:eq4}
\frac{dF}{p_Tdp_Td\eta d\phi}
= \left[\frac{dF}{p_{Ti}dp_{Ti}dy_id\phi_i} \frac{E}{E_i} \right]_{\mathbf{p}_i=\mathbf{p}-\mathbf{q}} \times\sqrt{1-\frac{m^2}{\left(m^2+p_T^2\right){\cosh}^2y}}.
\end{eqnarray}
In this equation, $dF/p_{Ti}dp_{Ti}dy_id\phi_i$ represents the normalized initial parton momentum distribution.
The factor $E/E_i$ ensures Lorentz invariance, and the conversion of rapidity to pseudorapidity is represented by $\sqrt{ \left. 1- m^2 \middle/ \left(m^2+p_T^2\right){\cosh}^2y \right.}$.
Here, $\mathbf{p}_i$ denotes the initial parton momentum, and $\mathbf{p}$ implies the final parton momentum shifted by $\mathbf{q}$.
Although $\mathbf{q}$ can vary with the jet particle’s momentum, we treat it as an average constant parameter that represents the statistically averaged momentum kick. 
The initial parton distribution represents the momentum distribution before collisions.
As two beams travel in opposite directions with equal energy, the net momentum of the two colliding particles in the beam direction is zero.
We assume that the jet particles come out in the perpendicular direction of the beam axis, and that the transverse momentum of the initial particles can be expressed as:
\begin{equation} \label{equation:eq5}
p_{Ti}^2=p_{T}^2-2p_{T} q \cos\left(\Delta\phi\right)+q^2.
\end{equation}

The initial parton momentum distribution is calculated using the following soft scattering model \cite{Wong_3, Wong_4, Wong_5}:
\begin{equation} \label{equation:eq6}
\frac{dF}{p_{Ti}dp_{Ti}dy_id\phi_i} =A_{\text{ridge}}\left(1-x\right)^a\frac{e^{ \left. -\sqrt{m^2+p_{Ti}^2} \middle/ T \right. }}{\sqrt{m_d^2+p_{Ti}^2}},
\end{equation}
where $A_{\text{ridge}}$ serves as the normalization constant for the initial partons, and $T$ represents the temperature of the medium particles.
Although the medium is composed of a variety of partons, we set the medium parton mass ($m$) to the pion mass because pions dominate the observed hadron composition. 
The fall-off parameter ($a$) determines the slope of the decrease in the $1-x$ distribution.
We examined the fall-off parameter with several values (0.05, 0.5, 1, and 5), and no significant differences were observed.
Consequently, we set $a = 0.5$, consistent with previous studies \cite{PbPb, Wong_1}, ensuring the general applicability of the MKM.

\begin{figure}
  \centering
  \includegraphics[width=10cm]{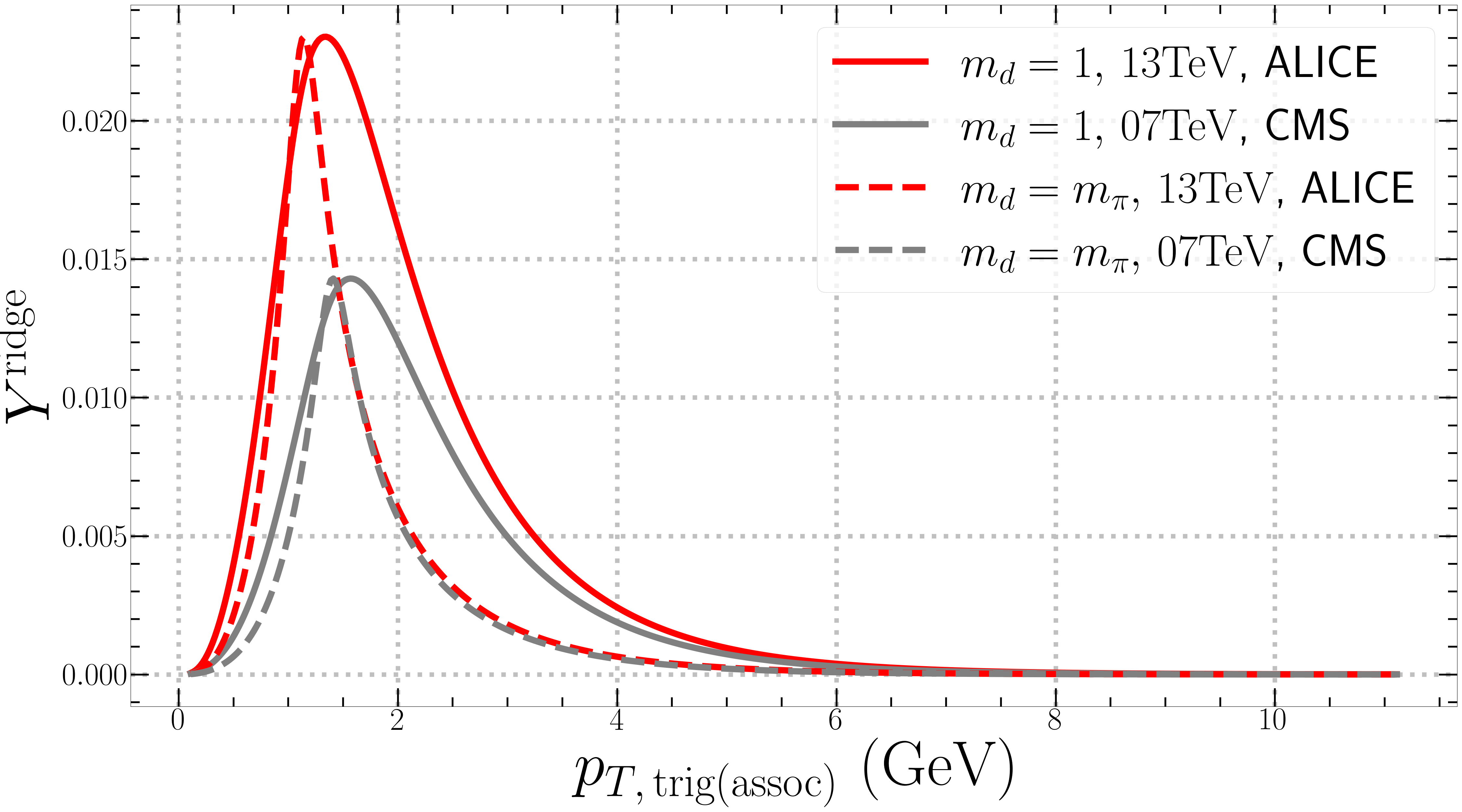}
  \caption{\label{figure:pTdis}
    Comparison of the ridge yield ($Y_{\text{ridge}}$) as a function of transverse momentum ($p_{T,\text{trig(assoc)}}$) for different $m_d$ values.
    Solid lines show $m_d = 1 \, \text{GeV}$, and dashed lines indicate $m_d = m_\pi$.
    The curves are normalized to the same peak value for shape comparison.
    The colors represent different experimental setups, with red indicating ALICE at 13 TeV and gray for CMS at 7 TeV.
  }
\end{figure}

In determining the cut-off parameter $m_d$, which prevents divergence at small $p_{Ti}$, we initially considered setting it to the pion mass ($m_\pi$) since pions are the most abundantly produced particles in collision experiments.
In contrast, previous studies set it to 1 GeV \cite{PbPb, Wong_1}, so we compared the ridge yield for $m_d = m_\pi$ and $1$ GeV, while keeping other parameters unchanged.
This results are shown in Figure \ref{figure:pTdis}
Here, the curves have been normalized to the same peak value for comparison.
The solid lines represent $m_d=1$ GeV, and the dashed lines represent $m_d=m_\pi$ in pp collisions at 13 TeV from ALICE (red) and 7 TeV from CMS (grey).

In the case of pp collisions at 13 TeV from ALICE (red), $m_d = m_\pi$ resulted in an overly narrow distribution with the Full Width at Half Maximum (FWHM) of approximately $0.7$ GeV, which failed to adequately describe the data, particularly in the $p_T > 1$ GeV region.
Setting $m_d=1$ GeV resulted in a FWHM of approximately $1.6$ GeV and provided a better fit to the observed data across a broader $p_T$ range.
A similar trend was observed in the CMS 7 TeV data (grey), indicating that $m_d = 1$ GeV offers a better fit across different energies.
Consequently, we used $m_d=1$ GeV in this study to explain various $p_T$ ranges, consistent with previous studies \cite{PbPb, Wong_1, Wong_2, Wong_3, Wong_4, Wong_5, Hanul}.
Similar trends were observed for 13 TeV data from other collaborations.

In Eq. (\ref{equation:eq6}), $x$ represents the light-cone variable defined as follows:
\begin{equation} \label{equation:eq8}
x=\frac{\sqrt{m^2+p_{Ti}^2}}{m_N}e^{\left|y_i\right|-y_b}.
\end{equation}
An increase in the daughter particle’s rapidity $( y_i $ leads to an exponential rise in the value of $x$.
This exponential behavior is determining the fraction of the parent’s longitudinal momentum that the daughter particle inherits during the production process \cite{Wong_Book}.
In the Eq.(\ref{equation:eq8}), $y_b$ corresponds to the beam rapidity, defined as $y_b = \cosh^{-1}{\sqrt{s_{\text{NN}}}/2m_N}$, with $m_N$ representing the mass of beam particles.
Since we aim to elucidate the ridge structure in proton-proton collisions, we set $m_N$ to the proton mass.

\begin{figure}
\centering
\includegraphics[width=12cm]{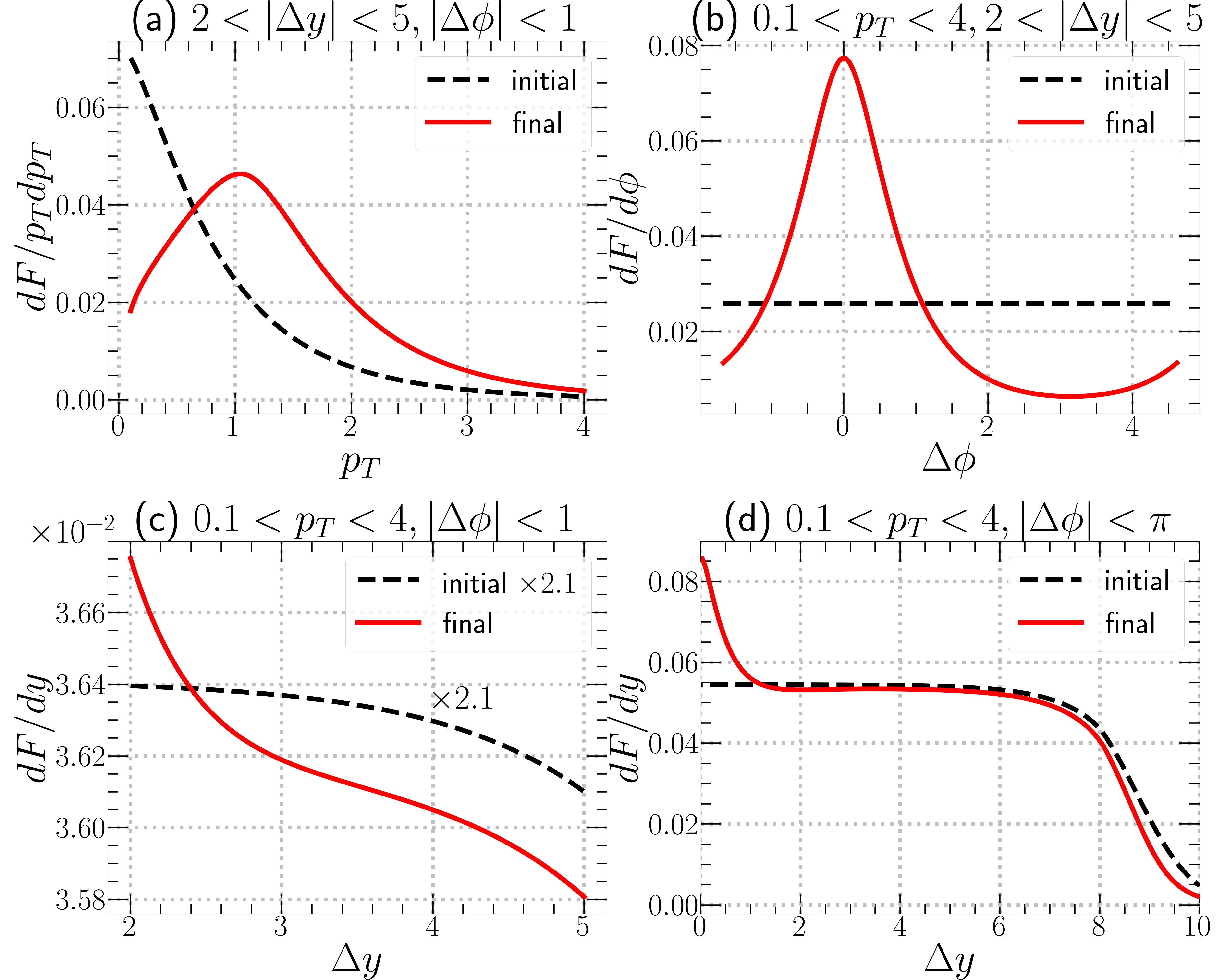}
\caption{
Comparison of the initial and final momentum distributions.
Panels (a), (b), and (c) show near-side distributions, while panel (d) covers the full azimuthal range.
The plotted quantities are $p_T$ (a), $\Delta \phi$ (b), and $\Delta y$ (c, d).
Black dashed curves correspond to initial partons, while red solid curves correspond to final partons.
In panel (c), the initial distribution is multiplied by 2.1 for clarity.
}
\label{figure:initialvsfinal}
\end{figure}

We investigate the jet-induced changes in the medium parton momentum distribution by comparing the initial and final partons.
The final parton momentum distribution is obtained without $f_R$ and $\langle N_k \rangle$.
Figure \ref{figure:initialvsfinal} represents how the distributions differ between the initial and final partons.
Panel (a) illustrates the $p_T$ distribution in the long-range ($2<|\Delta y|<5$) and near-side ($|\Delta \phi|<1$) regions.
We can observe that most medium partons initially have low transverse momentum.
However, after being kicked by the jet fragments, their $p_T$ distribution is shifted upward by approximately $q$.
Panel (b) presents the long-range $\Delta \phi$ distribution, revealing that medium partons, equally populated along $\phi$ initially, elongate toward the jet direction and converge near $\Delta \phi\sim0$.
Panel (c) focuses on the near-side $\Delta y$ distribution, indicating that partons with larger $\phi$ values are realigned toward smaller $\phi$, resulting in a significant increase in the final parton count in the near-side region.
For clearer comparison, the initial curve is multiplied by $2.1$.
Panel (d) displays the $\Delta y$ distribution over the full azimuthal range.
A distinct clustering of medium partons emerges near $\Delta y\sim0$, whereas the distribution away from $\Delta y \sim 0$ remains nearly unchanged.
This behavior reflects that the jet pseudorapidity ($\eta_{\text{jet}}=0$) makes the jet-induced kicks along the $y$ direction relatively small.
These results illustrate how jet-induced kicks can significantly reshape the transverse momentum, angular and rapidity distributions of medium partons.

$f_R \langle N_k \rangle$, $q$, and $T$ are the major parameters in the MKM to explain the ridge structure in the dihadron $\Delta\eta - \Delta\phi$ correlation.
Consequently, we will compare the parameter values with previous studies in Subsection \ref{subsec:pp_collisions_Fitting_results}.

\newpage
\section{\label{sec:FITTING_RESULTS}ANALYSIS}

\subsection{\label{subsec:Analysis_Procedure}Data Analysis in Experiments}

\begin{table*}[b]
    \setlength{\tabcolsep}{0.3cm}
    \renewcommand{\arraystretch}{1.3}
    \centering
    \begin{tabular}{c|ccc} \hline\hline
        & ALICE \cite{alice} & CMS \cite{cms} & ATLAS \cite{atlas} \\ \hline
         $\Delta \eta $ range & $1.6<|\Delta \eta |<1.8$ & $2<|\Delta \eta |<4$ & $2<|\Delta \eta |<5$\\
         high-multiplicity range& $0$--$0.1\%$ & $N_{\text{trk}}^{\text{offline}} \geq (105, 110)$& $N_{\text{ch}}^{\text{rec}} \geq 90$  \\
         $p_T$ range& $1<p_T<4$ & $0.1<p_T<4$ & $0.5<p_T<5$ \\ \hline\hline
    \end{tabular}
    \caption{\label{table:range}
      The experimental conditions for ALICE, CMS, and ATLAS \cite{alice, cms, atlas}.
    The high-multiplicity range of the pp collisions in the CMS is specified $N_{\text{trk}}^{\text{offline}}\geq 110$ for 13 TeV, and $N_{\text{trk}}^{\text{offline}}\geq 105$ for 7 TeV.
    In the case of ALICE, the high-multiplicity range at 13 TeV is selected based on the top $0.1\%$ of the total data.
        For ATLAS, the high-multiplicity range of pp collisions at 13 TeV is defined as $N_{\text{ch}}^{\text{rec}} \geq 90$ \cite{atlas}, and the data are available for different multiplicity ranges.
Additionally, the ALICE and CMS provide data for several $p_T$ bins, whereas the ATLAS groups all data into a whole set with no separate bins.
    }
\end{table*}

Experimentally, a two-particle correlation yield represents the ratio of particle pairs from the same events to mixed events, which is defined as follows \cite{alice}: 
\begin{equation}\label{correlation yield}
\frac{1}{N_{\text{trig}}}\frac{d^2N^{\text{pair}}}{d\Delta\eta d\Delta\phi} =B(0,0)\frac{S(\Delta\eta, \Delta\phi)}{B(\Delta\eta, \Delta\phi)} \Bigg|_{p_{T}^{\text{trig}}, \ p_{T}^{\text{assoc}}},
\end{equation}
where $\Delta\eta$ and $\Delta\phi$ are the differences in $\eta$ and $\phi$ between two randomly selected particles.
The numerator, $S(\Delta\eta, \ \Delta\phi)$, denotes the per-trigger-particle yield of particle pairs originating from the same event.
The denominator, $B(\Delta\eta, \Delta\phi)$, represents the mixed-event background, and $B(0,0)$ serves to normalize $B(\Delta\eta, \Delta\phi)$ \cite{alice}.

We aim to describe the long-range near-side $\Delta \phi$ correlation data from three major collaborations conducted by the ALICE, CMS, and ATLAS, altogether \cite{alice, cms, atlas}.
However, since the experimental conditions of these three are different, it is necessary to check them, which are summarized in Table \ref{table:range}.
The ALICE covers a narrower $\Delta \eta$ range compared to that of the CMS and ATLAS due to its detector setup.
Additionally, the definition of high-multiplicity events differs among the collaborations.
The ALICE selects the top 0.1\% of all events \cite{alice}, the CMS defines them based on the number of tracks \cite{cms}, and the ATLAS defines them based on the number of reconstructed charged particles \cite{atlas}.
Furthermore, both the ALICE and CMS provide data in several $p_T$ bins, whereas the ATLAS analyzes data over a broader $p_T$ range from 0.5 to 5 GeV as a whole.

In addition, there are differences in analysis methods.
The ALICE and CMS utilize the ZYAM (Zero Yield At Minimum) method, which involves subtracting the minimum value of the data from the entire dataset to set the minimum value to zero.
The ZYAM method assumes that the jet contribution does not appear at the minimum $\Delta \phi$ correlation value \cite{zyam}, which is a traditional approach in experimental analysis.
In this method, $C_{\text{ZYAM}}$ means the minimum yield at $\Delta \phi = \Delta \phi_{\text{ZYAM}}$.
There is a slight difference in how $C_{\text{ZYAM}}$ is defined between the ALICE and CMS, but this difference is negligible \cite{alice, cms}.

However, the ATLAS utilized a peripheral subtraction method, in which the peripheral component (low multiplicity) is subtracted from the high-multiplicity data to isolate the pure high-multiplicity contribution \cite{atlas}.
To eliminate uncertainty in the experimental analysis, we restored the original data and applied the ZYAM method.
The resulting values of $C_{\text{ZYAM}}$ and $\Delta \phi_{\text{ZYAM}}$ are presented in Table \ref{table:CZYAM}, which we utilized.
\begin{table}
\setlength{\tabcolsep}{1cm}
\renewcommand{\arraystretch}{1.3}
\centering
\begin{tabular}{c|c|c}  \hline\hline
$N_{\text{ch}}^{\text{rec}}$ & $\Delta\phi_{\text{ZYAM}}$ & $C_{\text{ZYAM}}$ \\ \hline
$90 \leq N_{\text{ch}}^{\text{rec}}$        & 0.96 & 5.47 \\ \hline
$50 \leq N_{\text{ch}}^{\text{rec}} < 60$   & 0.61 & 2.88 \\ \hline
$60 \leq N_{\text{ch}}^{\text{rec}} < 70$   & 0.79 & 3.44 \\ \hline
$70 \leq N_{\text{ch}}^{\text{rec}} < 80$   & 0.96 & 3.99 \\ \hline
$80 \leq N_{\text{ch}}^{\text{rec}} < 90$   & 0.96 & 4.55 \\ \hline
$90 \leq N_{\text{ch}}^{\text{rec}} < 100$  & 0.96 & 5.10 \\ \hline
$100 \leq N_{\text{ch}}^{\text{rec}} < 110$ & 0.96 & 5.65 \\ \hline
$110 \leq N_{\text{ch}}^{\text{rec}} < 120$ & 0.96 & 6.19 \\ \hline
$120 \leq N_{\text{ch}}^{\text{rec}} < 130$ & 0.96 & 6.73 \\ \hline
$130 \leq N_{\text{ch}}^{\text{rec}}$       & -0.96 & 7.57 \\ \hline\hline
\end{tabular}%
    \caption{\label{table:CZYAM}
    Values of $\Delta \phi_{\text{ZYAM}}$ and $C_{\text{ZYAM}}$ used in this study to adjust the ATLAS $\Delta \phi$ correlation data.
    }
\end{table}

\newpage
\subsection{\label{subsec:pp_collisions_Fitting_results}Fitting results for high-multiplicity pp collisions at 13 and 7 TeV}

Given that recent research on ridge structures in small systems has primarily focused on the high-multiplicity region, we begin our analysis by applying the MKM to high-multiplicity pp collisions at 13 TeV and 7 TeV.
The major parameters of the MKM for describing the ridge structure in two-particle correlations are $q$, $T$, and $f_R \langle N_k \rangle$.
Jet particles scatter off medium partons several times while they pass through the medium. Whenever they scatter, they transfer the momentum to the medium partons.
Even though the number of scatterings per jet particle is fewer in pp collisions than in AA collisions, jet particles transfer the momentum to the medium partons. $q$ is the average value of momentum transfer per kick over several scattering events and over the jet fragment particles.
The temperature, $T$, denotes the temperature of medium partons, which is more like an average kinetic energy rather than a statistical one.
$f_R$ and $\langle N_k \rangle$ are the ratio of surviving particles at the detector and the average number of collisions with medium particles, respectively, and they work as a single parameter such as $f_R \langle N_k \rangle$.

Table \ref{table:param} compares the values of the major parameters in the MKM, obtained from our analysis of high-multiplicity pp collisions at 13 TeV and 7 TeV, alongside previous results for high-multiplicity pp collisions at 13 TeV \cite{Hanul}.
The second column presents the results from the previous study for pp collisions at 13 TeV, while the third and fourth columns show the results from our current analysis.
Let us examine the major parameter values in greater detail, addressing each individually.
\begin{table}[b]
\centering
\setlength{\tabcolsep}{0.4cm}
\renewcommand{\arraystretch}{1.1}
\begin{tabular}{c|c|c|c} \hline\hline
                            & Ref. \cite{Hanul} &            \multicolumn{2}{c}{This study}\\ \hline
                            & pp 13 TeV        & pp 13 TeV  & pp 7 TeV \\
                            & CMS, ATLAS       & ALICE, CMS & CMS \\ \hline
                  $T$ (GeV) & $1.54$           & $1.19$     & $1.17$ \\
                  $q$ (GeV) & $2$              & $1.12$     & $1.39$ \\
$f_R \langle N_k \rangle$ & $0.93$--$1.37$& $2.13$ & $1.05$ \\ \hline\hline
\end{tabular}%
\caption{\label{table:param}
The physical parameters for the MKM.
Previous findings from Ref. \cite{Hanul} on pp collisions at 13 TeV reveal varying $f_R \langle N_k \rangle$ across $p_T$ bins.
In the previous study (second column), the temperature was fixed using the $\langle p_T \rangle$ ratio from AuAu collisions at 200 GeV, whereas, in our analysis (third and fourth columns), it was treated as a free parameter.
}
\end{table}

First, regarding the temperature $T$, we can observe a significant difference even at the same center-of-mass energy of 13 TeV.
We treated the temperature as a free parameter for the 13 TeV data from the ALICE and CMS, unlike the previous study that fixed it based on the $\langle p_T \rangle$ ratio from AuAu collisions at 200 GeV \cite{Wong_1, Hanul}.
This adjustment was necessary because pp collisions at 13 TeV represent a small-system environment with significantly higher center-of-mass energy than AuAu at 200 GeV, 65 times higher energy, which makes the previous method less reliable.
In addition, the abundance of data makes it possible to treat $T$ as a free parameter.
The observed increase of $T$ in pp collisions relative to Au+Au can be attributed to the larger total momentum of the measured particles \cite{Hanul}. 
Although we similarly observe a rapid increase in $T$, the ratio $T/T_{\text{AuAu 200GeV}}$  falls from 3.08 to 2.38, reflecting the substantially different experimental environment compared to heavy-ion collisions \cite{Hanul}.
Since this methodological difference also influences other parameters, a comparison between our results and those in Ref. \cite{Hanul} is not meaningful.
Therefore, we will limit our comparison to 13 TeV (3rd column) and 7 TeV (4th column) results.
Although the temperature for 13 TeV was treated as a free parameter, the temperature for 7 TeV can be fixed using the $\langle p_T \rangle$ ratio:
\begin{equation} \label{equation:temp7tev} 
T_{7 \ \text{TeV}} = T_{13 \ \text{TeV}} \times \frac{\langle p_T \rangle_{7 \ \text{TeV}}}{\langle p_T \rangle_{13 \ \text{TeV}}} = 1.17,
\end{equation}
because, unlike those in Ref. \cite{Hanul}, the experimental environments are nearly identical for both center-of-mass energies.
This assumption can be reasonably understood since temperature reflects the statistical average of kinetic motion.
We used the $\langle p_T \rangle$ values from the ATLAS \cite{ATLAS:2010jvh, ATLAS:2016zkp} in Eq. (\ref{equation:temp7tev}), which $\langle p_T \rangle _{7 \ \text{TeV}} = 1.226$ GeV and $\langle p_T \rangle _{13 \ \text{TeV}} = 1.245$ GeV, matching the corresponding multiplicity ranges to CMS data \cite{cms}, due to the availability of experimental data.
We expect that different temperatures would result from different center-of-mass energies, but limited experimental data prevent a comprehensive study, requiring further research.

Second, we explore the average momentum transfer per kick, $q$.
Contrary to our expectation that this parameter is proportional to center-of-mass energy, Table \ref{table:param} shows that $q$ increases from $1.12$ GeV at 13 TeV to $1.39$ GeV at 7 TeV as the energy decreases.
Thus, it is necessary to examine the relationship between center-of-mass energy and $q$.
Ref. \cite{ATLAS:2016zkp} shows that multiplicities are generally proportional to the center-of-mass energy.
Since higher multiplicity leads to more frequent collisions for each jet particle, we expect an inverse proportional relationship between multiplicity and $q$.
To check this inverse proportionality, we calculated the product of the $q$ value ratio and the average multiplicity ratio between 13 TeV and 7 TeV, using the multiplicity values from Ref. \cite{ATLAS:2016zkp}:
\begin{equation} \label{equation:KickMultiRatio}
\frac{q_{\text{13 TeV}}}{q_{\text{7 TeV}}} \times \frac{\langle N_{\text{ch}}^{\text{13 TeV}} \rangle}{\langle N_{\text{ch}}^{\text{7 TeV}}\rangle} = \frac{1.12}{1.39} \times \frac{120.4}{104.9} \approx 0.93.
\end{equation}
This product is close to 1, supporting their inverse proportionality.
$q$ and multiplicity are anti-proportional, and center-of-mass energy and multiplicity are proportional to each other.
Therefore $q$ and center-of-mass energy are anti-proportional.
We will use this relationship when incorporating the multiplicity dependence into the model in Subsect \ref{subsec:How_about_the_other_multiplicity?}.
However, additional research is required to understand the underlying reasons for this relationship fully, including in other collision systems.

Third, for the parameter $f_R \langle N_k \rangle$, $p_T$ dependence is implemented in some studies \cite{PbPb, Hanul}, while it is not in others \cite{Wong_1, Wong_2, Wong_3, Wong_4, Wong_5}.
During our fitting process for high-multiplicity pp collisions at 13 TeV, we observed that the $p_T$ dependence naturally disappeared.
Therefore, we also set $f_R \langle N_k \rangle$ as a constant for 7 TeV, given the similarity of the environments in pp collisions at both energies.

Once $f_R \langle N_k \rangle$ is set to be a constant, it serves as an overall amplitude in addition to the original contribution from the survival factor and the number of kicks partons.
This explains the smaller $f_R \langle N_k \rangle$ at 7 TeV compared to 13 TeV in Table \ref{table:param}.

\begin{figure*}
\centering
\includegraphics[width=\linewidth]{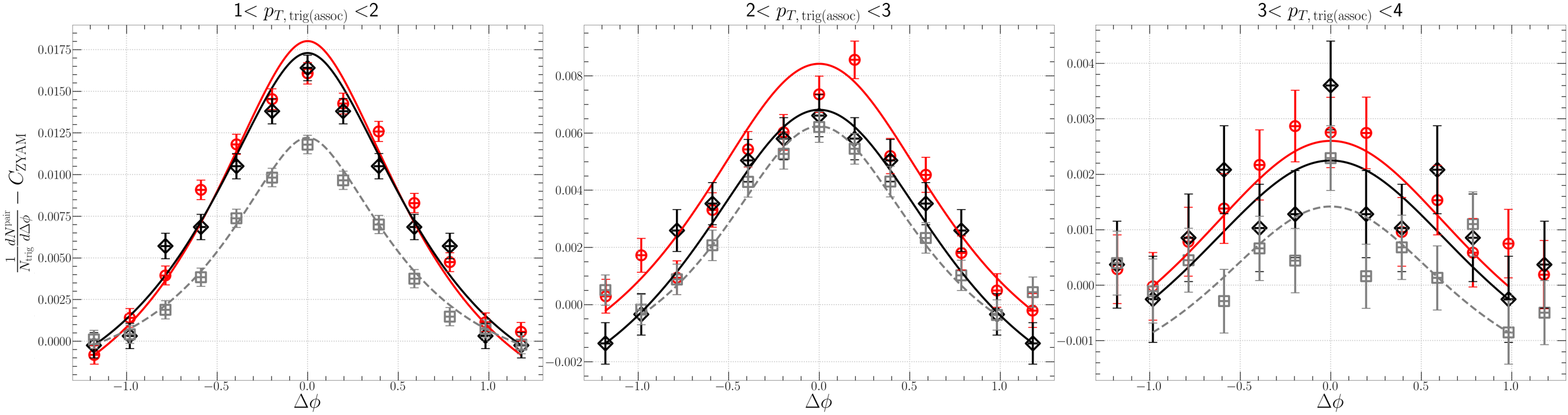}
\caption{
Fitting results for dihadron $\Delta \phi$ correlations in $1<p_T<2$, $2<p_T<3$, and $3<p_T<4$ GeV ranges.
All curves are the MKM results, and all symbols are experimental data.
Moreover, the circles represent the ALICE data (red), the diamonds denote the CMS 13 TeV data (black), and the squares describe the CMS 7 TeV data (grey).
}
\label{figure:phicorr}
\end{figure*}

Figure \ref{figure:phicorr} shows the MKM results for the $\Delta \phi$ correlation fitted to the high-multiplicity pp collision data from the ALICE \cite{alice} and CMS \cite{cms}.
The solid lines represent the MKM results, and the symbols indicate the experimental data from ALICE and CMS.
The circles and diamonds correspond to the 13 TeV pp collision data from ALICE (red) and CMS (black), respectively, while the squares show the 7 TeV pp collision data from CMS (grey).
We can see that the MKM describes the experimental data well for the available $p_T$ ranges in both ALICE and CMS.

\begin{figure}
  \centering
  \includegraphics[width=10cm]{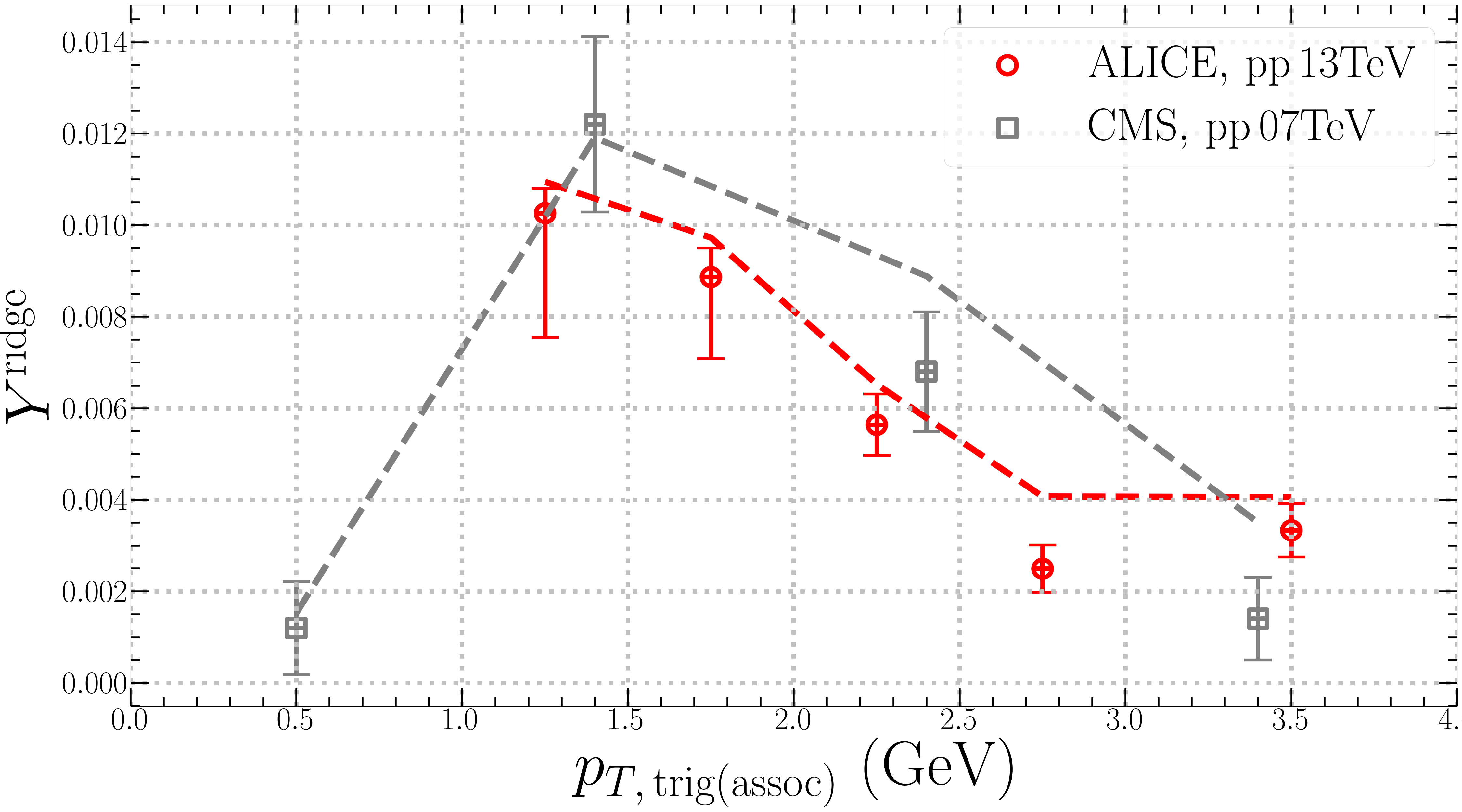}
  \caption{
    The MKM results for $Y^{\text{ridge}}$.
Red squares and grey squares represent experimental data from ALICE pp at 13 TeV and CMS pp at 7 TeV, respectively.
Dashed lines connect the corresponding theoretical results. 
    The $p_T$ bins used for the theoretical results are chosen to match those of the experimental data.
  }
  \label{figure:Yridge_scatter}
\end{figure}

Figure \ref{figure:Yridge_scatter} presents the MKM results for the yield.
Grey circles and red squares represent experimental data from CMS pp at 7 TeV \cite{cms} and ALICE pp at 13 TeV \cite{alice}, respectively.
Theoretical results are connected by dashed lines.

\begin{figure}
\centering
\includegraphics[width=9cm]{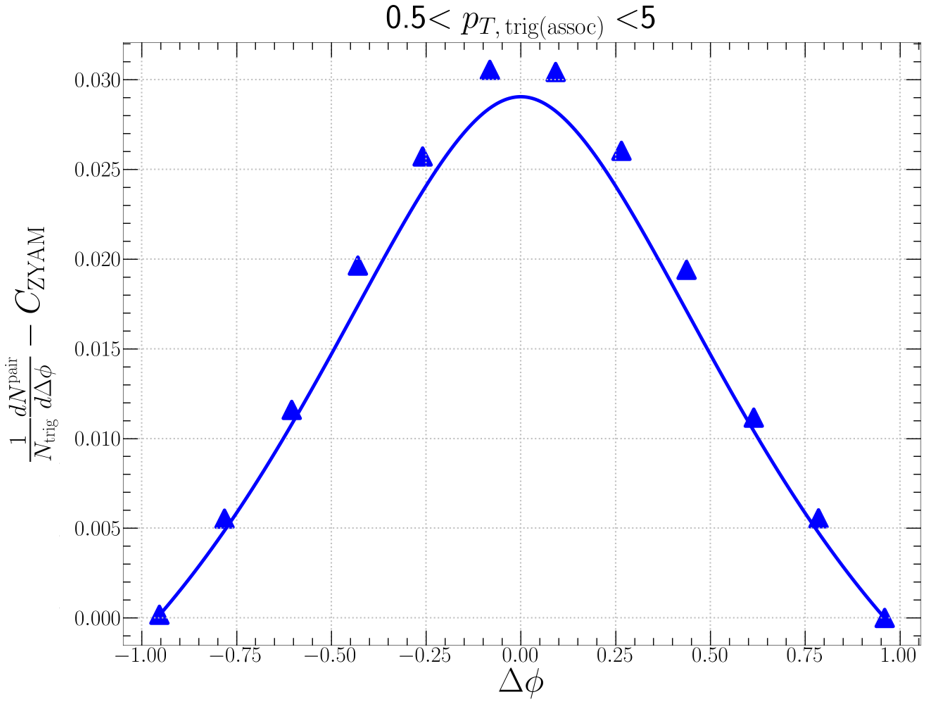}
\caption{\label{figure:phicorr_atlas}
The MKM result of dihadron $\Delta \phi$ correlation for the ATLAS experiment for $0.5<p_T<5$ GeV range.
The solid line represents the MKM result.
This result is calculated using parameter values obtained from fitting the ALICE and CMS data and adjusted to match the ATLAS experimental conditions.
The symbols represent the ATLAS data.
}
\end{figure}
However, we wonder if these key parameters, $q$, $T$, and $f_R \langle N_k \rangle$ in the MKM can be applied to different experimental environments.
Thus, we tested this application on the ATLAS pp collision at 13 TeV data, which covers a slightly broader $p_T$ range of $0.5<p_T<5 \, \text{GeV}$ compared to ALICE and CMS.
Figure \ref{figure:phicorr_atlas} shows the MKM results applied to the ATLAS experimental conditions, using the parameter values obtained from fitting the ALICE and CMS data.
The MKM results appear to underestimate the peak region of the ATLAS data.
However, as the detailed experimental uncertainties are unavailable, it is expected that the MKM results may adequately describe the data within those uncertainties.
Therefore, due to this broad applicability, these parameter values can serve as a reference for applying the MKM to other experiments.

\subsection{\label{subsec:How_about_the_other_multiplicity?}Extension to Other Multiplicity Ranges}

We observed that the MKM can adequately explain the ridge structure in high-multiplicity pp collisions at 13 TeV and 7 TeV.
Since the MKM did not originally include the multiplicity effects, we hope to enhance the model by incorporating the multiplicity dependence.
Among the available $\Delta \phi$ correlation data for various multiplicities from the CMS and ATLAS, we chose the ATLAS data due to its superior precision \cite{atlas, ATLAS:2016zkp}.
We introduced the multiplicity dependence into the major parameters $q$, $T$, and $f_R \langle N_k \rangle$.

First, to set the parameter $q$, we use its inverse proportional relationship to the multiplicity, as found in Subsect \ref{subsec:pp_collisions_Fitting_results}:
\begin{equation} \label{equation:variousmulti_kick}
  q = q_{\text{high}} \times \frac{\langle N_{\text{ch}}^{\text{rec}} \rangle_{\text{high}}}{\langle N_{\text{ch}}^{\text{rec}} \rangle},
\end{equation}
where `high' means high-multiplicity, and $\langle N_{\text{ch}}^{\text{rec}} \rangle$ is the mean value of the specific multiplicity bin.

Second, since the temperature reflects the statistical average of kinetic motion, we adjust it based on the $\langle p_T \rangle$ ratio for each multiplicity bin, using the method applied in Subsect \ref{subsec:pp_collisions_Fitting_results}.
This can be expressed using the following equation:
\begin{equation} \label{equation:variousmulti_temp}
  T = T_{\text{high}} \times \frac{\langle p_T \rangle}{\langle p_T \rangle_{\text{high}}}.
\end{equation}
Since $\langle p_T \rangle$ is proportional to multiplicity, this implies that the temperature increases with multiplicity.

\begin{figure}
  \centering
  \includegraphics[width=10cm]{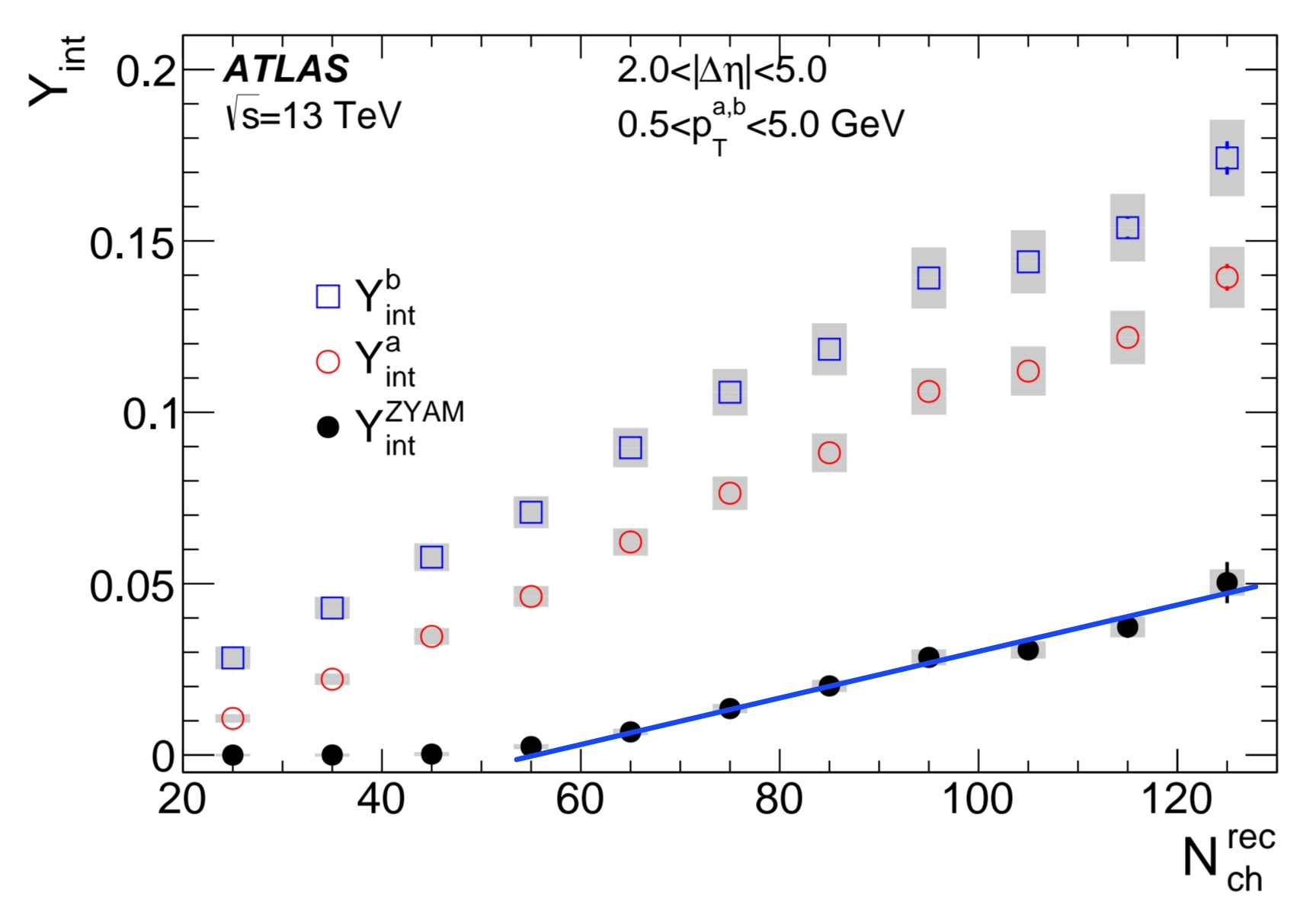}
  \caption{\label{figure:frnk_multi}
  Integrated long-range near-side yield versus multiplicity in proton-proton collisions at 13 TeV.
  The plots denote the results from Ref. \cite{atlas}, and we only utilize black plots.
  The blue line represents the linearly fitted line on black plots, and we only consider the range of $50 \leq N_{\text{ch}}^{\text{rec}}$.
  The fitting result is $(-35+0.65\langle N_{\text{ch}}^{\text{rec}} \rangle)\times 10^{-3}$, which is used for $f_R \langle N_k \rangle$ in the $\Delta \phi$ correlation fitting.
  }
\end{figure}

Third, since $f_R \langle N_k \rangle$ reflects the amplitude when treated as a constant, it should capture the trend of the overall yield.
Figure \ref{figure:frnk_multi} represents the integrated near-side yields as a function of multiplicity \cite{atlas}.
\begin{table}
    \setlength{\tabcolsep}{0.3cm}
    \renewcommand{\arraystretch}{1.5}
    \centering
    \begin{tabular}{ c  |c|c|c |c  |c | c} \hline \hline
        $\langle N_{\text{ch}}^{\text{rec}}\rangle$& $N$ range &$\langle p_T \rangle$&  $q$   &   $T$   & $A$ &$f_R \langle N_k \rangle$\\ \hline\hline
        $108.83$& $88.5$--$150.5$& $1.228$& $1.12$& $1.19$& &$2.13$\\ \hline\hline
        $54.5$& $49.5$--$60.5$&$1.123$& $2.237$& $1.086$& $0.082$&$0.118$\\ \hline 
        $63.58$& $59.5$--$70.5$&$1.133$& $1.917$& $1.105$& $0.265$ &$0.306$\\ \hline 
        $73.48$& $68.5$--$80.5$&$1.165$& $1.659$& $1.125$& $0.523$ &$0.571$\\ \hline 
        $83.47$& $78.5$--$90.5$&$1.185$& $1.460$& $1.144$& $0.926$ &$0.980$\\ \hline\hline
        $93.38$& $88.5$--$100.5$&$1.200$& $1.305$& $1.160$& $1.604$ &$1.665$\\ \hline 
        $104.8$& $100.5$--$110.5$&$1.216$& $1.163$& $1.178$& $1.953$ &$2.021$\\ \hline 
        $114.8$& $110.5$--$120.5$&$1.241$& $1.062$& $1.203$& $3.202$ &$3.277$\\ \hline 
        $124.7$& $120.5$--$130.5$&$1.259$& $0.977$& $1.220$& $5.062$ &$5.144$\\ \hline 
        $137.4$& $130.5$--$150.5$&$1.263$& $0.887$& $1.224$& $5.949$ &$6.038$\\ \hline\hline

    \end{tabular}
    \caption{\label{table:variousmulti}
    The results of the physical parameters and the values used for each multiplicity range.
    The first column represents the average multiplicity of the multiplicity range shown in the second column.
The sixth column represents the overall constant `$A$' in $f_R \langle N_k \rangle$, and the last column represents $f_R \langle N_k \rangle$ itself.
  The second row shows parameters found in the MKM using an overall high-multiplicity range fitted for ALICE and CMS in Subsect \ref{subsec:pp_collisions_Fitting_results}.
  $q$ and $T$ values in various multiplicity bins are calculated from Eq. (\ref{equation:variousmulti_kick})--(\ref{equation:variousmulti_frnk}) based on these values.
  The last five rows represent the high-multiplicity part defined by the ATLAS \cite{atlas}.
} 
\end{table}
The $y$-axis ($Y_{\text{int}}$) represents the integrated near-side yield, while the $x$-axis ($N_{\text{ch}}^{\text{rec}}$) signifies the multiplicity.
The black dots denote $\displaystyle Y_{\text{int}}^{\text{ZYAM}}$, the integrated near-side yield using the ZYAM method.
We observed that the overall yield of near-side particles increases with the multiplicity.
Consequently, we introduced linear multiplicity dependence into $f_R \langle N_k \rangle$, following the blue solid line in Figure \ref{figure:frnk_multi}, which is
\begin{equation} \label{equation:variousmulti_frnk}
  f_R\langle N_k \rangle = A+0.00065 \langle N_{\text{ch}}^{\text{rec}} \rangle.
\end{equation}
In Eq. (\ref{equation:variousmulti_frnk}), the parameter `$A$' represents an overall constant of each multiplicity bin, which is the only free parameter when fitting the $\Delta \phi$ correlation data.

To determine the parameter values in Eq. (\ref{equation:variousmulti_kick})--(\ref{equation:variousmulti_frnk}), we need a dataset that provides $\langle p_T \rangle$ values for various multiplicity bins.
It was obtained from Ref. \cite{ATLAS:2016zkp}, which provides more detailed segmentation than Ref. \cite{atlas}.
Thus, we combined some bins and averaged $\langle p_T \rangle$ values weighted by multiplicity.
The detailed values and fitting results can be found in Table \ref{table:variousmulti}.

\begin{figure*}
\centering
\includegraphics[width=\linewidth]{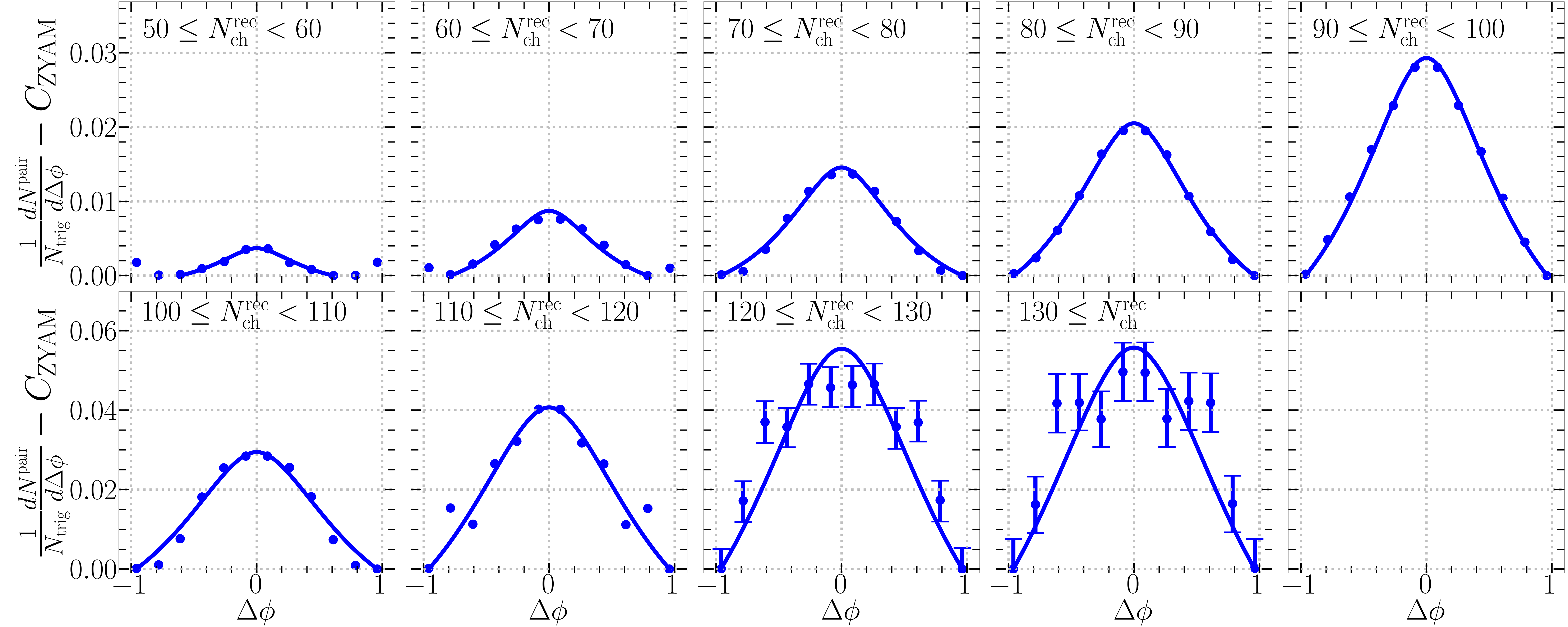}
\caption{$\Delta \phi$ correlation results for various multiplicity ranges at $0.5 < p_T < 5$ GeV.
The symbols show the ATLAS experimental data and the solid lines represent the MKM results.
}
\label{figure:variousmulti}
\end{figure*}

As can be seen in Table \ref{table:variousmulti}, it is evident that $q$ decreases and $T$ increases as the multiplicity increases.
This is due to the relationships introduced in Eq. (\ref{equation:variousmulti_kick}) and (\ref{equation:variousmulti_temp}), respectively.
Additionally, we can observe an increase in the overall constant `$A$'.
Although we incorporated a linear multiplicity dependence into the parameter $f_R \langle N_k \rangle$, the general increase in the overall constant suggests a nonlinear relationship between $f_R \langle N_k \rangle$ and the multiplicity.
Further investigation is required to determine the precise multiplicity dependence of $f_R \langle N_k \rangle$.
Figure \ref{figure:variousmulti} shows the results of the MKM and the ATLAS $\Delta \phi$ correlation data. 
Overall, the MKM can describe the $\Delta\phi$ correlation reasonably well for various multiplicity ranges.

\clearpage
\section{\label{sec:Prediction}Prediction}

The MKM has effectively described the dihadron $\Delta\phi$ correlations in pp collisions at 13 TeV and 7 TeV.
With the LHC Run 3 currently underway, featuring slightly higher center-of-mass energy than Run 2, we provide predictions for high-multiplicity pp collisions at 14 TeV using the MKM.

\begin{figure*}[h]
  \centering
  \includegraphics[width=\linewidth]{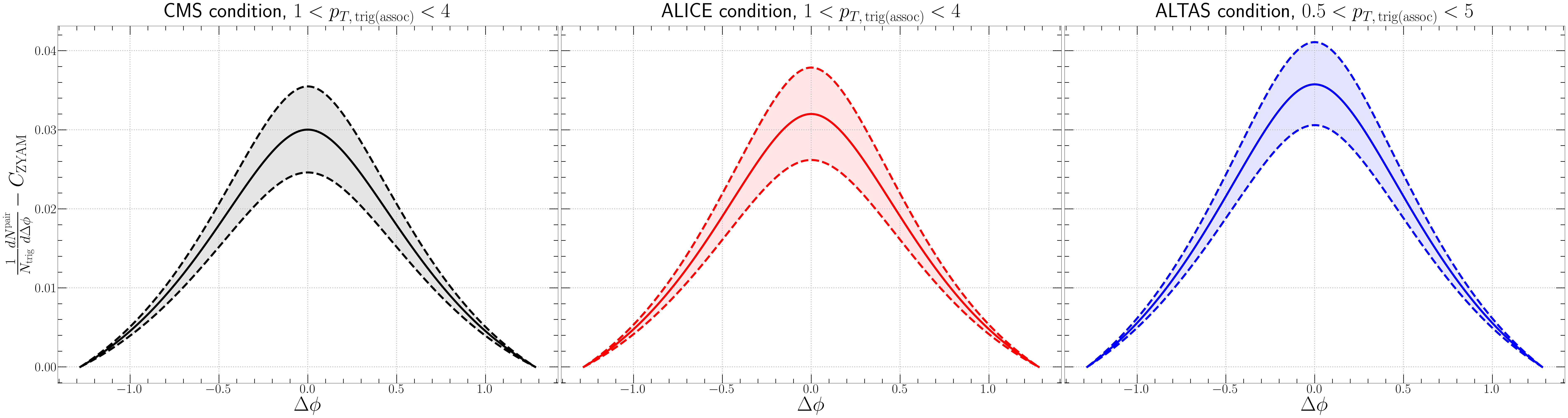}
  \caption{
Predictions of the $\Delta \phi$ correlation at 14 TeV from the MKM for CMS (black), ALICE (red), and ATLAS (blue).
The ALICE and CMS predictions are plotted for $1<p_T<4$ GeV, while ATLAS is plotted for $0.5<p_T<5$ GeV.
These predictions are based on the parameters extrapolated from the 7 TeV and 13 TeV results in Subsect \ref{subsec:pp_collisions_Fitting_results}.
The parameter values are adjusted by $\pm 10\%$ to determine upper and lower bounds.
}
  \label{figure:pp14prediction}
\end{figure*}

At this collision energy, we do not have the multiplicity and $\langle p_T \rangle$ values needed to predict the key parameters of the MKM.
Therefore, we apply the energy ratio between high-multiplicity pp collisions at 13 TeV and 7 TeV, as obtained from Subsection \ref{subsec:pp_collisions_Fitting_results}, to extend our predictions to 14 TeV.
This approach is based on the assumption that the multiplicity and $\langle p_T \rangle$ are proportional to the center-of-mass energy.
We adjust parameter values by $\pm 10\%$ to derive the upper and lower bounds.
Table \ref{tab:PredictionTable} shows the predicted parameter values for pp collisions at 14 TeV, and Figure \ref{figure:pp14prediction} illustrates the MKM's prediction for the $\Delta \phi$ correlation in high-multiplicity pp collisions at 14 TeV.
Three plots represent the CMS (black), ALICE (red), and ATLAS (blue) experimental conditions, respectively.

\begin{table}[h]
    \setlength{\tabcolsep}{0.8cm}
    \renewcommand{\arraystretch}{1.2}
    \centering
    \begin{tabular}{c | ccc}
        \hline\hline
         Parameters&  $-10\%$&  Middle& $+10\%$\\ \hline
         $T$&  $1.074$&  $1.193$& $1.312$\\ 
         $q$&  $0.968$&  $1.075$& $1.183$\\
         $f_R\langle N_k \rangle$&  $2.079$&  $2.310$& $2.541$\\ \hline\hline
    \end{tabular}%
    \caption{\label{tab:PredictionTable}
    Parameters predicting the 14 TeV result.
    The parameters are derived by extending the energy ratio from high-multiplicity 13 TeV and 7 TeV, with adjustments of $\pm 10\%$ applied to determine the upper and lower bounds.
}
\end{table}

\section{\label{sec:Conclusion}CONCLUSIONS}

As the near-side ridge structure was discovered in high-multiplicity small systems, several theories have been developed to explain it.
In this study, we applied the Momentum-Kick Model (MKM) to proton-proton collisions at $\sqrt{s}  = 13$ TeV and $\sqrt{s}  = 7$ TeV, focusing on the long-range near-side region.
We utilized data from the ALICE, CMS, and ATLAS for 13 TeV and data from the CMS for 7 TeV \cite{alice, cms, atlas}.

In the case of high-multiplicity pp collisions at 13 TeV, we began by fitting the ALICE and CMS data.
In the prior study \cite{Hanul}, the temperature was fixed using the $\langle p_T \rangle$ ratio from AuAu collisions at 200 GeV, whereas in this study it was treated as a free parameter.
This is because the collision environment in pp collisions at 13 TeV is too different from that in AuAu at 200 GeV, making the previous method unreliable.
The 13 TeV dataset is sufficiently large, allowing us to treat a sufficient number of model parameters as free during the fitting process.
We observed that the $p_T$ dependence of the parameter $f_R\langle N_k \rangle$, introduced in previous studies \cite{PbPb, Hanul}, naturally vanishes.
The fitting parameters are found to be $T=1.19$ GeV, $q=1.12$ GeV, and $f_R\left\langle N_k \right\rangle = 2.13$ for pp collisions at 13 TeV.
We then applied these values to the ATLAS data, which covers a wider transverse momentum($p_T$) range of $0.5$--$5$ GeV.
Remarkably, the MKM successfully described the ATLAS data, demonstrating the model's robustness across different experimental conditions and $p_T$ ranges.

In the case of high-multiplicity pp collisions at 7 TeV, we set the temperature using the $\langle p_T \rangle$ ratio, since the temperature reflects the statistical average of kinetic motion.
In addition, $f_R \langle N_k \rangle$ is set to be constant as in the 13 TeV.
The fitting parameters are found to be $T=1.17$ GeV, $q=1.39$ GeV, and $f_R\left\langle N_k \right\rangle = 1.05$, with $q$ observed to be larger at 7 TeV than at 13 TeV.
This is verified by the inverse proportionality between $q$ and center-of-mass energy.
For $f_R \langle N_k \rangle$, the 7 TeV results are lower than those at 13 TeV due to the smaller yield at 7 TeV.

To account for the $\Delta \phi$ correlation across various multiplicity ranges in pp collisions at 13 TeV, we introduced a multiplicity dependence in the MKM.
We set $q$ to be inversely proportional to multiplicity and determined $T$ using the $\langle p_T \rangle$ ratio corresponding to each multiplicity bin.
In this process, the parameter values that served as the base are from the high-multiplicity 13 TeV data.
Since $f_R \langle N_k \rangle$ represents the total number of particles in the model, we introduced a linear dependence as a function of multiplicity.
During the fitting of the $\Delta \phi$ correlation data, we added an overall constant `$A$' to the $f_R\langle N_k \rangle$ to describe the yields for each multiplicity bin.
Consequently, we found $q=2.237$, $T=1.086$, and $A=0.221$ for $50\leq N_{\text{ch}}^{\text{rec}}<60$ (the lowest multiplicity range), and $q=0.887$, $T=1.224$, and $A=0.111$ for $130\leq N_{\text{ch}}^{\text{rec}}$ (the highest multiplicity range).
The overall constant `$A$' increases across multiplicities, despite introducing a linear multiplicity dependence into $f_R \langle N_k \rangle$.
It may imply additional multiplicity dependence in $f_R \langle N_k \rangle$, which requires further study.

In conclusion, the MKM is a very simplified model with few parameters through scattering but has successfully described the ridge structure in small systems. It is expected to be applicable to future LHC results.
To predict the pp collision at 14 TeV data, we extrapolated parameter values from results at 7 TeV and 13 TeV.
In order to establish the upper and lower bounds, we adjusted parameter values by $\pm 10\%$.
This prediction result can be tested through future experiments in the LHC.

Having demonstrated the efficacy of the MKM in elucidating the ridge structure observed in pp collisions at 13 TeV and 7 TeV, it is necessary to ascertain whether the model can also account for other center-of-mass energies and asymmetric systems.
In addition, more detailed studies are required on the relationship of the parameters, $q$ and $T$, to the center-of-mass energy and multiplicity, as well as the behavior of $f_R \langle N_k \rangle$, especially its $p_T$ dependence.

\section*{Acknowledgements}
This research is supported by the Inha University Research Fund.

\bibliographystyle{elsarticle-harv} 
\bibliography{main}

\end{document}